\UseRawInputEncoding
\documentclass[aps,prx,superscriptaddress,reprint]{revtex4-1}
\usepackage[usenames,dvipsnames]{xcolor}
\usepackage{amsmath}
\usepackage{amssymb}
\usepackage{graphicx}
\usepackage{url}
\usepackage{hyperref}
\usepackage{color,xcolor}
\usepackage{epstopdf}
\usepackage{float}
\usepackage{ulem}
\usepackage[percent]{overpic}
\usepackage{siunitx}
\usepackage{subdepth}
\usepackage{enumitem}

\usepackage{dcolumn}% Align table columns on decimal point
\usepackage{bm}% bold math
\usepackage{times}
\usepackage[version=3]{mhchem} % Formula subscripts using \ce{}

\usepackage{mathrsfs}
\usepackage{lipsum}
\usepackage{amsfonts}
\usepackage{array}
\usepackage{ulem}
\usepackage{lineno}
\usepackage[none]{hyphenat}
\hypersetup{colorlinks=true,citecolor=Red,linkcolor=Blue,urlcolor=Black}
\begin{document}
\title{Simulation of Topological $X$-Gates via Braiding of Majorana Zero Modes in an Interacting Quantum Dot System}

\author{Bradraj Pandey*}
\affiliation{Department of Physics and Astronomy, The University of 
Tennessee, Knoxville, Tennessee 37996, USA}
\affiliation{Materials Science and Technology Division, Oak Ridge National 
Laboratory, Oak Ridge, Tennessee 37831, USA}
\affiliation{Department of Physics and Astronomy, University of Missouri, Columbia, Missouri 65211, USA}

\author{Satoshi Okamoto}
\affiliation{Materials Science and Technology Division, Oak Ridge National 
Laboratory, Oak Ridge, Tennessee 37831, USA}

\author{Pontus Laurell}
\affiliation{Department of Physics and Astronomy, University of Missouri, Columbia, Missouri 65211, USA}
\affiliation{Materials Science and Engineering Institute, University of Missouri, Columbia, Missouri 65211, USA}

\author{Stephan Rachel}
\affiliation{ School of Physics, University of Melbourne, Parkville, VIC, Australia}

\author{Elbio Dagotto}
\affiliation{Department of Physics and Astronomy, The University of 
Tennessee, Knoxville, Tennessee 37996, USA}
\affiliation{Materials Science and Technology Division, Oak Ridge National 
Laboratory, Oak Ridge, Tennessee 37831, USA}

\date{\today}
\begin{abstract}
Recent advances in quantum dot platforms have opened new platform for realizing Majorana zero modes (MZMs) and simulating topological quantum computation. 
Here we propose an experimentally feasible setup for implementing topological $\sqrt{X}$- and $X$-quantum gates in an interacting $Y$-shaped quantum-dot array. 
The proposed novel architecture enables both braiding and charge readout through simple fusion operations controlled by gate-tunable potentials.
Using many-body time-dependent simulations based on exact diagonalization, we analyze the braiding and fusion dynamics of MZMs in the presence of nearest-neighbor 
Coulomb interactions and pairing disorder. We compute diabatic errors, braiding fidelity, and the time- and space-resolved electron and hole components of 
the local density of states to monitor the braiding process. Our results show that even weak interactions or pairing disorder induce oscillations 
	in the braiding fidelity, thereby setting an upper bound on the braiding speed. Furthermore, we demonstrate that comparing fusion outcomes 
	before and after braiding provides a direct and experimentally accessible signature of the non-Abelian nature of MZMs in quantum dot systems.
\end{abstract}
\maketitle

\noindent {\bf Introduction} \\
Majorana zero modes (MZMs) are charge-neutral quasiparticle excitations that emerge as subgap zero-energy states 
in topological superconductors~\cite{Kitaev1,Kitaev2}. Their non-Abelian exchange statistics and nonlocal encoding of quantum 
information make them promising candidates for fault-tolerant topological quantum computation~\cite{Kitaev2,Sarma,Nayak,Shnirman}.
Over the past two decades, the most promising experimental platforms for realizing MZMs have been
engineered heterostructures utilizing superconducting proximity effect: the two most investigated platforms being
semiconductor–superconductor hybrid structures~\cite{Sau,Oreg2010} and ferromagnetic spin
chains deposited on conventional superconducting substrates~\cite{Yazdani}.
However, these systems face several challenges, 
including the formation and detection of MZMs in the presence of disorder and impurities.
Recent advances in quantum-dot-based systems provide a promising alternative platform, 
offering significantly reduced disorder and enhanced tunability through electrostatic gate control~\cite{Jay,Dvir,Bordin,tenHaaf}. 
In particular, experiments using two or three quantum dots coupled via a short superconducting–semiconducting hybrid segment 
(e.g., InSb nanowires) have demonstrated the realization of short Kitaev chains~\cite{Dvir,Bordin}. 
Notably, in these experiments, two spatially separated MZMs were detected via tunneling conductance measurements at the sweet spot, 
where the interdot hopping amplitude and superconducting order parameter are equal, i.e. \( t^h = \Delta \).

Quantum-dot arrays offer a promising  platform for simulating topological quantum computation
and testing non-Abelian statistics through controlled exchange (braiding) of MZMs~\cite{Boross2024,Tsintzis2024,Miles2025}. 
In quantum-dot systems, Majorana zero modes (MZMs) are localized on-site at the sweet spot~\cite{Dvir,Bordin},
enabling fusion and braiding even in comparatively small system sizes, in contrast to conventional nanowire-based platforms.
The experimental realization of MZMs braiding can provide definitive evidence of non-Abelian statistics and, 
importantly, distinguish them from unwanted trivial Andreev bound states.
Probing such non-Abelian properties requires the adiabatic exchange of MZMs using minimal geometries such as
\(T\)- or \(Y\)-shaped quantum-dot networks~\cite{Boross2024,Tsintzis2024,Pandey2025,Pandey2023,Sekania2017}. 
The controlled motion of MZMs has been demonstrated experimentally in three-site quantum dot
configurations~\cite{tenHaaf}, and numerically in six-dot systems~\cite{Pandey2025_PRB}.
Realistic modeling of such braiding protocols demands full many-body, 
time-dependent simulations that incorporate the effects of both the
degenerate ground states and bulk excitations. 
While prior studies on proximitized nanowire systems often relied on non-interacting~\cite{Harper,Amorim2015} 
or low-energy effective models, braiding of four MZMs in next-generation quantum-dot architectures remains less explored~\cite{Sekania2017,Miles2025}. 

\begin{figure}[!ht]
\hspace*{-0.5cm}
\vspace*{0cm}
\begin{overpic}[width=1.0\columnwidth]{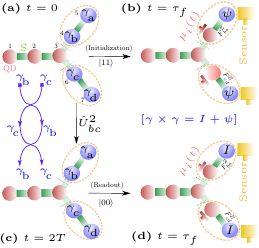}
\end{overpic}
\caption{%
{\bf Initialization, braiding, and readout of a Majorana zero-mode qubit in a $Y$-shaped quantum-dot array.} 
Schematic of a device comprising seven quantum dots coupled via semiconductor–superconductor hybrid structures arranged in a $Y$-shaped geometry. 
Red dots indicate non-topological sites ($\mu_j = \mu_0$), while blue dots denote topological sites ($\mu_j = 0$). 
\textbf{a}, At $t = 0$, two pairs of Majorana zero modes (MZMs), $(\gamma_a, \gamma_b)$ and $(\gamma_c, \gamma_d)$, are localized at the ends of the second and third legs, respectively. 
\textbf{b}, $Z$-fusion operations, $F^Z_{ba}$ and $F^Z_{cd}$, are performed by tuning the on-site chemical potentials $\mu_j(t)$ at sites $j=4$ and $6$. 
The fusion outcome is an electron $\psi$ on sites $j=5$ and $6$. The outcomes are detected through charge sensing by nearby quantum-dot sensors (yellow) coupled to sites $j=5$ and $7$. 
\textbf{c}, Illustration of the double-braiding operation $\hat{U}_{bc}^2$, exchanging MZMs $\gamma_b$ and $\gamma_c$. 
\textbf{d}, Final readout is performed via $Z$-fusion of $(\gamma_a, \gamma_b)$ and $(\gamma_c, \gamma_d)$ after the double-braiding process. 
The final fusion outcome is the vacuum $I$ on sites $j=5$ and $6$, which differs from the initial fusion outcomes.%
}
\label{fig1}
\end{figure}

In this work, we propose a novel scheme to test non-Abelian statistics in an interacting quantum-dot system arranged in
a $Y$-shaped geometry (Fig.~\ref{fig1}). Our experimentally feasible setup, composed of only a few quantum dots, 
enables the initialization, braiding, and readout of a MZM-based qubit within its coherence lifetime~\cite{Bordin}.
We perform dynamical simulations of an $X$-gate in an interacting 
quantum-dot system using the full many-body wavefunction. We simulate the braiding of MZMs in interacting 
quantum dots arranged in a $Y$-shaped geometry by solving the time-dependent many-body Schrödinger 
equation via exact diagonalization. 
To assess the fidelity of the braiding protocol, we compute the diabatic error and transition probabilities.
We visualize the positions of the MZMs during the $X$-gate process by calculating the time-, space-,
and energy-resolved local density of states, $LDOS(t,j,\omega)$~\cite{Pandey2023_PRB,Thomale2013},
which should be accessible in time-dependent differential conductance measurements using
scanning tunneling spectroscopy~\cite{Houselt2010}.

Moreover, we show that performing fusion of the same MZM pairs (Z-fusion) before and after braiding provides a 
simple and experimentally accessible readout mechanism for the qubit state in quantum-dot systems.
Specifically, we evaluate the time-dependent electron and hole components of the 
LDOS($t,j,\omega$) to resolve the fusion outcome and track the MZMs dynamics during braiding. 
We also compute the time-dependent local charge density, which can be directly measured using 
 charge-sensing techniques in quantum-dot experiments~\cite{Szechenyi2020,Gharavi2016}.
Furthermore, we investigate the robustness of the braiding protocol against key experimental imperfections,
including nearest-neighbor Coulomb repulsion, charge noise, and spatial disorder in the pairing amplitude. 
We find that nearest-neighbor Coulomb repulsion and pairing disorder lead to the hybridization of MZMs, 
which in turn gives rise to Majorana oscillations. This effect imposes an upper bound on the braiding speed in the presence of interactions.
Our numerical results indicate that initialization, braiding, and readout of MZMs can be successfully implemented in 
quantum-dot platforms using our proposed 
setup, well within the typical quasiparticle poisoning timescales observed in current experiments~\cite{Bordin,tenHaaf}.

\noindent {\bf \\Results\\}

\noindent {\small \bf Model Hamiltonian\\}
Next-generation quantum-dot experimental platforms are composed of spin-polarized quantum dots 
coupled via short superconductor–semiconductor hybrid segments~\cite{Dvir,Bordin}. The inter-dot coupling is mediated by Andreev bound states 
through elastic cotunneling (ECT) and crossed Andreev reflection (CAR), enabling gate-tunable single-particle hopping \( t^h \) 
and effective triplet pairing \( \Delta \)~\cite{Bordin2023}. After fine-tuning the system to the so-called sweet spot, 
where \( t^h = \Delta \), localized Majorana zero modes (MZMs) emerge at the edges of the chain~\cite{Dvir,Bordin}.

Here, we introduce a minimal and experimentally feasible setup consisting of {\it seven} interacting quantum dots arranged in a $Y$-shaped geometry (see Fig.~\ref{fig1}),
suitable for implementing MZM braiding operations and readout. In this configuration, the quantum dots are connected via electrostatically tunable gates, 
which enable dynamic control over the hopping amplitude \( t^h(t) \), pairing strength \( \Delta(t) \), and on-site chemical potential 
\( \mu_j(t) \)~\cite{Dvir,Bordin,Liu2025,Pandey2024_PRR}, each carrying a time dependence. This high degree of tunability allows for the adiabatic transport of MZMs from one 
quantum dot to another~\cite{tenHaaf,Pandey2025_PRB}, thereby enabling both braiding and fusion operations of MZMs.
The Hamiltonian for the $Y$-shaped quantum dot array (with $\phi_{I} = 0$, $\phi_{II} = \pi/6$, and $\phi_{III} = -\pi/6$ for each leg~\cite{Harper})    
    can be divided into four distinct parts. The Hamiltonian for each arm is given by~\cite{Sekania2017}:                                                

\begin{eqnarray}
H^I &=& \sum_{j=1}^{l}\left( -t^h c^{\dagger}_{j} c_{j+1} + \Delta c_j c_{j+1} + \text{H.c.} \right) + \sum_{j=1}^{l+1} \mu_j(t)\, n_j, \nonumber \\
	H^{II} &=& \sum_{j=l+2}^{2l}\left( -t^h c^{\dagger}_{j} c_{j+1} + e^{i\phi_{II}} \Delta c_j c_{j+1} + \text{H.c.} \right), \nonumber \\    
	H^{III} &=& \sum_{j=2l+2}^{3l}\left( -t^h c^{\dagger}_{j} c_{j+1} + e^{i\phi_{III}} \Delta c_j c_{j+1} + \text{H.c.} \right)
\label{eq:H123}
\end{eqnarray}
    Additionally, the Hamiltonian for the central site $l+1$, which connects the second and third arms, is:                           
\begin{align}
	H^{IV} &= -t^h_{l+1, l+2}(t)\, c^{\dagger}_{l+1} c_{l+2} 
	+ e^{i\phi_{II}} \Delta_{l+1 l+2}(t)\, c_{l+1} c_{l+2} + \text{H.c.} \nonumber \\
	&\quad -t_{l+1 2l+2}(t)\, c^{\dagger}_{l+1} c_{2l+2} 
	+ e^{i\phi_{III}} \Delta_{l+1,2l+2}(t)\, c_{l+1} c_{2l+2} + \text{H.c.}
\label{eq:H4}
\end{align}

\begin{figure}[h]
\centering
\includegraphics[scale=1.92]{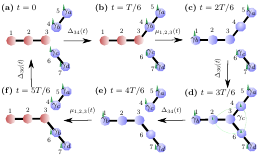}
\caption{{\bf Braiding protocol for double exchange of Majorana zero modes in a $Y$-shaped quantum-dot array.} 
\textbf{a}, At time $t=0$, two spatially separated pairs of MZMs, $(\gamma_a, \gamma_b)$ and $(\gamma_c, \gamma_d)$, are localized at sites $j=4,5$ and $j=6,7$, 
  respectively. The chemical potentials $\mu_{1,2,3}$ are set to $\mu_0$, and the hopping and pairing amplitudes between the second and 
	third arms are initially switched off, i.e., $\Delta_{34}(t) = \Delta_{36}(t)= t^h_{34}(t)=t^h_{36}(t)= 0$. 
	\textbf{b}, The coupling $\Delta_{34}(t)=t^h_{34}(t)$ between sites $j=3$ and $j=4$ is smoothly ramped up using a sine-squared protocol. 
\textbf{c}, Simultaneously, the chemical potentials $\mu_{1,2,3}$ are reduced from $\mu_0$ to zero, transferring $\gamma_b$ from site $j=4$ to $j=1$ at $t=2T/6$. 
	\textbf{d}, The coupling $\Delta_{36}(t)=t^h_{36}(t)$ is then turned on, delocalizing $\gamma_c$ across sites $j=2$, $j=4$, and $j=6$. 
	\textbf{e}, The coupling $\Delta_{34}(t)=t^h_{34}(t)$ is gradually turned off, localizing $\gamma_c$ back at site $j=4$ at $t=4T/6$. 
\textbf{f}, Finally, the chemical potentials $\mu_{1,2,3}$ are restored to $\mu_0$, moving $\gamma_b$ from site $j=1$ to $j=6$. 
	The coupling $\Delta_{36}(t)=t^h_{36}(t)$ is subsequently turned off, completing a single braid  ($t=T$) where $\gamma_b$ and $\gamma_c$ have been exchanged. 
  }
\label{fig2}
\end{figure}

%\begin{eqnarray}
%	H^{IV} &=& \left( -t^h(t)\, c^{\dagger}_{l+1} c_{l+2} + e^{i\phi_{II}} \Delta(t)\, c_{l+1} c_{l+2} + \text{H.c.} \right) \nonumber\\
%	&+& \left( -t^h(t)\, c^{\dagger}_{l+1} c_{2l+2} + e^{i\phi_{III}} \Delta(t)\, c_{l+1} c_{2l+2} + \text{H.c.} \right)
%\label{eq:H4}
%\end{eqnarray}
Here, $l = (L-1)/3$, where $L$ is the total number of sites, $c_j$ is the fermionic annihilation operator, and $n_j = c^{\dagger}_{j} c_j$ is the
number operator. The parameters $t^h_{ij}$ and $\Delta_{ij}(t)$ denote the bond-dependent hopping and pairing amplitudes, respectively.
     We assume $t^h=\Delta$ for our numerical calculations (unless stated otherwise). 
We employ a smooth time-dependent sine-squared ramp protocol, which has been shown to outperform the linear ramp protocol~\cite{Pandey2025_PRB,Sekania2017}. 
The protocol is defined as
\[
    \mu(t) = \mu_0 \sin^2\left(\frac{\pi t}{2\tau}\right), 
    \quad 
    t^h_{ij}(t) = \Delta_{ij}(t) = \Delta_0 \sin^2\left(\frac{\pi t}{2\tau}\right),
\]
where $\mu_0 = 12$, $\Delta_0 = 1.0$, and $\tau$ is the switching time (larger $\tau$ corresponds to slower gate tuning).

\noindent {\small \bf Braiding Protocol}

In topological quantum computation, a single Majorana qubit can be encoded using four Majorana zero modes (MZMs)~\cite{Sarma}.
These four MZMs form a four-fold degenerate ground state. Among them, two states with occupancy $|0,0\rangle$ and $|1,1\rangle$, 
share the same even fermion parity, while the other two, $|0,1\rangle$ and $|1,0\rangle$, have odd fermion parity. 
The exchange (braiding) of MZMs induces a unitary rotation within the manifold of 
degenerate ground states with fixed total parity~\cite{Nayak}. 
These states constitute the computational basis of an MZM-based qubit, 
and braiding operations correspond to quantum gates acting on this subspace~\cite{Sarma,Marra2022}. 
In our proposed setup, which uses a $Y$-shaped geometry composed of only seven quantum dots, 
the X-gate is realized by the double exchange of two central MZMs, $\gamma_b$ and $\gamma_c$.

At the initial time \( t = 0 \), the system hosts two spatially separated pairs of Majorana zero modes (MZMs):
\((\gamma_a, \gamma_b)\) located at sites \( j = 4, 5 \), and \((\gamma_c, \gamma_d)\) at sites \( j = 6, 7 \). 
The chemical potentials \(\mu_{1,2,3}\) are initialized at a high value \(\mu_0=12\), 
while the hopping and pairing terms connecting the second and third arms are set to zero, i.e., \(\Delta_{34}(t)=t^h_{34}(t) = 0\) and \(\Delta_{36}(t) =t^h_{36}(t)= 0\).
Initialization of the system can be achieved by performing simple Z-fusion operations between MZMs within each pair~\cite{Hodge2025,Malciu2018}.
This is implemented through the fusion operations \( F^Z_{ba} \) and \( F^Z_{cd} \), 
realized by increasing the on-site chemical potential at sites \( j=4,6 \) adjacent to the fusion sites \( j=5,7 \)
( see Fig.~\ref{fig1}). After initialization, the braiding sequence is carried out through the following steps 
(see Fig.~\ref{fig2}):

\begin{enumerate}[label=(\roman*)]

\item The hopping and pairing terms between sites \( j = 3 \) and \( j = 4 \) are smoothly ramped from 0 to 1 using
a sine-squared protocol over a time interval \( T/6 \), thereby turning on \(\Delta_{34}(t)=t^h_{34}(t)\).

\item The chemical potentials \(\mu_{1,2,3}\) are simultaneously decreased from \(\mu_0\) to 0 via a sine-squared ramp,
causing the MZM \(\gamma_b\) to move from site \( j = 4 \) to site \( j = 1 \) by time \( t = 2T/6 \).

\item The coupling \(\Delta_{36}(t)=t^h_{36}(t)\) between sites \( j = 3 \) and \( j = 6 \) is turned on,
leading to the formation of a multi-site MZM \(\gamma_c\) delocalized across sites \( j = 2, 4, 6 \) at time \( t = 3T/6 \).

\item The coupling \(\Delta_{34}(t)=t^h_{34}(t)\) between sites \( j = 3 \) and \( j = 4 \) is reduced from 1 to 0,
relocalizing the MZM \(\gamma_c\) at site \( j = 4 \) by time \( t = 4T/6 \).

\item The chemical potentials \(\mu_{1,2,3}\) are increased from 0 back to \(\mu_0\),
driving \(\gamma_b\) from site \( j = 1 \) to site \( j = 6 \) by time \( t = 5T/6 \).

\item Finally, the coupling \(\Delta_{36}(t)=t^h_{36}(t)\) between sites \( j = 3 \) and \( j = 6 \) is smoothly turned off,
completing a single braid between \(\gamma_b\) (now at \( j = 6 \)) and \(\gamma_c\) (now at \( j = 4 \)) at time \( t = T \).

\end{enumerate}

Repeating the same sequence results in a double braid at time \( t = 2T \), returning \(\gamma_b\) and \(\gamma_c\) 
to their original positions as at \( t = 0 \). In the ideal adiabatic limit, a single braid is described by the unitary operator  
\( U_{bc} = \exp\left(\frac{\pi}{4} \gamma_b \gamma_c\right) \), whereas a double braid corresponds to \( U_{bc}^2 = \gamma_b \gamma_c \) (see also in SM). 
The double braid operation transforms the initial state into an orthogonal one within the same fermion-parity sector of 
the degenerate ground-state manifold, which is a signature of non-Abelian statistics. Notably, in quantum-dot experiments, 
the double braid protocol offers a clearer demonstration of non-Abelian behavior 
compared to a single braid~\cite{Tsintzis2024,Clarke2017}. For this reason, 
we focus primarily on the double braid operation throughout this work.
The final readout of the system is performed by repeating the same Z-fusion operations 
\( F^Z_{ba} \) and \( F^Z_{cd} \) on each MZM pair after the braiding sequence.\\

\noindent {\small \bf $\sqrt{X}$ and $X$-gate simulation} 

\begin{figure*}[!ht]
\hspace*{-0.5cm}
\vspace*{0cm}
\begin{overpic}[width=2.1\columnwidth]{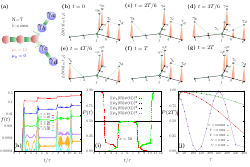}
\end{overpic}
\caption{{\bf Simulation of Majorana braiding and qubit dynamics in a $Y$-shaped quantum-dot array.}
\textbf{a}, Initial configuration at $t=0$ with two pairs of spatially separated Majorana zero modes (MZMs), $(\gamma_a, \gamma_b)$ and $(\gamma_c, \gamma_d)$.
\textbf{b--g}, Space- and time-resolved local density of states $LDOS(\omega, t, j)$ at different stages of the braiding sequence. 
\textbf{b}, At $t=0$, four zero-energy peaks are observed on sites $j=4,5,6,7$, corresponding to the four MZMs.
\textbf{c}, At $t=2T/6$, the MZM $\gamma_b$ is transferred from site $j=4$ to $j=1$, leading to zero-energy peaks at $j=1,5,6,7$. 
\textbf{d}, At $t=3T/6$, $\gamma_c$ becomes partially delocalized, resulting in finite spectral weight at sites $j=3,4,6$, while sharp peaks remain at $j=1,5,6$. 
\textbf{e}, At $t=4T/6$, $\gamma_c$ becomes localized at site $j=4$, and zero-energy peaks are observed at $j=1,4,5,7$. 
\textbf{f}, At $t=T$, after completing a single braid, $\gamma_b$ and $\gamma_c$ are exchanged, as indicated by the restored peaks at $j=4,5,6,7$. 
\textbf{g}, At $t=2T$, following a full double braid, the system returns to its initial configuration, showing four zero-energy peaks identical to the initial state.
\textbf{h}, Time evolution of the diabatic error $f(t) = 1 - \sum_{n=1}^{4} |\langle \psi(t) | \psi_n^I \rangle |^2$ as a function of scaled time $t/\tau$ for different ramp durations $\tau$.
\textbf{i}, Time evolution of the transition probabilities $P_n(t) = |\langle \psi_n(0) | \psi(t) \rangle |^2$ 
  for all four eigenstates ($n=1,2,3,4$), shown for $\tau=50$.
\textbf{j}, Final transition probability $P(2T)$ at the end of the protocol ($t=2T$) as a function of switching time $\tau$ for different values of the Coulomb interaction strength $V$.}
\label{fig3}
\end{figure*}

Here, we focus on simulating the \(X\)-gate~\cite{Mascot2023,Bedow2024} using four Majorana zero modes (MZMs) in a 
seven-quantum-dot setup. We perform time evolution using the Hamiltonian described in Eqs.~\eqref{eq:H123} and \eqref{eq:H4}, 
starting from the initial state \( |\psi_1(0)\rangle \) with total even fermion number parity \( P_{\text{tot}} = +1 \).
As shown in Fig.~\ref{fig3}(a), at \( t = 0 \), two distinct pairs of MZMs, \((\gamma_a, \gamma_b)\) and \((\gamma_c, \gamma_d)\)
 are localized at sites \( j = 5,4 \) and \( j = 6,7 \), respectively. 
The \(X\)-gate corresponds to a double exchange of the two inner MZMs, \(\gamma_b\) and \(\gamma_c\), 
within the $Y$-shaped geometry.
At this initial time, the fermion parity of each MZM pair is found to be 
\( P_{ba} = -i \langle \psi_1(0)| \gamma_b \gamma_a |\psi_1(0)\rangle = -1 \) and \( P_{cd} = -i \langle \psi_1(0)| \gamma_c \gamma_d |\psi_1(0)\rangle = -1 \).
To monitor the braiding dynamics at each step, we compute the local density of states at zero energy, 
\( \text{LDOS}(\omega = 0, j, t) \), as a function of space and time (see Fig.~\ref{fig3}(b)–(g)). 
At \( t = 0 \), \( \text{LDOS}(\omega = 0, j, t) \) exhibits four equal-height peaks at sites \( j = 4,5,6,7 \), 
corresponding to four localized MZMs on the two separate legs of the $Y$-shaped quantum-dot array (Fig.~\ref{fig3}(b)).
Notably, at time \( t = 3T/6 \), \( \text{LDOS}(\omega = 0, j, t) \) shows three distinct peaks at sites \( j = 2, 4, 6 \),
indicating that \(\gamma_c\) becomes delocalized across these sites and forms a multi-site MZM (Fig.~\ref{fig3}(d)). 
This demonstrates that a minimum of seven quantum dots is required to prevent overlap between \(\gamma_c\) 
and other MZMs at the system boundaries during braiding.
At the intermediate time \( t = T \), \( \text{LDOS}(\omega = 0, j, t) \) again shows four peaks at sites 
\( j = 4,5,6,7 \), similar to the configuration at \( t = 0 \), but with \(\gamma_b\) and \(\gamma_c\) having exchanged positions, 
indicating the completion of a single braid operation (Fig.~\ref{fig3}(f)).
At the final time \( t = 2T \), \(\gamma_b\) and \(\gamma_c\) return to their original positions, 
confirming the successful execution of a full braiding sequence (Fig.~\ref{fig3}(g)).
The braiding process, as revealed by the evolution of \( \text{LDOS}(\omega, t, j) \), should be experimentally accessible
via time-resolved differential conductance measurements~\cite{Houselt2010}.

In Fig.~\ref{fig3}(h), we evaluate the diabatic error as a function of scaled time \( t/\tau \), defined by  
\[
f(t) = 1 - \sum_{n=1}^{4} \left| \langle \psi(t) | \psi_n^I \rangle \right|^2,
\]  
where \( \psi_n^I \) denotes the instantaneous many-body eigenstates of the system. 
For small values of \(\tau\) (corresponding to fast switching), 
the diabatic error takes a larger value, indicating transitions to higher-energy states above the gap. 
Interestingly, we find that for \(\tau \geq 40\), the diabatic error remains low (\( f(t) \sim 10^{-4} \)),
showing negligible excitation above the ground-state manifold.
For larger \(\tau\), the spikes in the diabatic error appear at times when the on-site potentials 
at the three initial sites $(j=1,2,3)$ are changed simultaneously. 
The diabatic error for smaller values of switching time leads to larger braiding error and gives a lower
bound to the switching time $\tau \gtrsim \hbar/\Delta_{gap}$~\cite{Knapp2016,Peeters2024} (here $\Delta_{gap}$ is the excitation gap in the topological phase). 

Figure~\ref{fig3}(i) shows the time evolution of the transition probabilities \( P_n(t) = |\langle \psi_n(0)|\psi(t)\rangle|^2 \) 
within the ground-state manifold (\(n = 1, 2, 3, 4\)) as a function of \( t/\tau \), with switching time \( \tau = 50 \). 
Initially, at \( t = 0 \), the system is entirely in the ground state, with \( P_1(0) = 1 \) and all other \( P_n(0) = 0 \).
Interestingly, in the intermediate time range \( 4 \lesssim t/\tau \lesssim 8 \), near the completion of a single braid, 
both \( P_1(t) \) and \( P_4(t) \) approach \( 1/2 \), indicating an equal superposition of two many-body states with 
the same total fermion parity. In the Majorana occupation basis, this corresponds to the action of a \(\sqrt{X}\)-gate: 
\( |11\rangle \rightarrow \frac{1}{\sqrt{2}}(|11\rangle + i |00\rangle) \)~\cite{Mascot2023}.
At \( t/\tau = 12 \), the probability \( P_4(t) \) approaches 1 while \( P_1(t) = 0 \), indicating successful completion of a double braiding 
operation that transforms \( |\psi_1(0)\rangle \rightarrow |\psi_4(0)\rangle \). In the Majorana occupation basis, 
this final state represents a full \( X \)-gate: \( |11\rangle \rightarrow |00\rangle \), 
corresponding to a qubit flip of the MZM-based logical state.

In Fig.~\ref{fig3}(j), we show the effect of nearest-neighbor Coulomb interactions \( V \) 
on the braiding fidelity \( P(2T) \), defined as the transition probability at the final time \( t = 2T \)~\cite{Peeters2024}, 
plotted as a function of the switching time \( \tau \). In quantum-dot experiments, nearest-neighbor Coulomb interactions can be suppressed by 
the screening effect provided by the superconducting segments separating the dots. 
For this reason, we restrict our analysis to small interaction strengths in the range $0 \leq V \leq 0.3$.
Our numerical results show that for very weak interaction strengths 
(\( V \lesssim 0.0001 \)), the braiding fidelity remains close to unity for \( \tau \gtrsim 40 \), 
even at large switching times (Fig.~\ref{fig3}(j)). As \( V \) increases, \( P(2T) \) gradually decreases and 
reaches lower values for longer switching durations.
For interaction strengths \( V \gtrsim 0.001 \), the fidelity exhibits pronounced oscillations 
as a function of switching time~\cite{Harper,Amorim2015} (see also Supplementary Material). 
This behavior arises because increasing \( V \) reduces the localization of Majorana zero modes (MZMs), 
causing them to hybridize and thereby lifting the degeneracy of the qubit subspace. 
The resulting ground-state energy splitting introduces a dynamical phase, 
which in turn gives rise to oscillations in the braiding fidelity~\cite{Harper,Amorim2015}.
For intermediate values of interaction strength,
we find that the oscillation frequency of the braiding fidelity can be accurately described by fitting
\( P(2T) = \cos^2\left( \frac{T \omega}{2} \right) \)~\cite{Peeters2024,Hodge2025PRL}.
These results indicate that, in the presence of interaction \( V \), braiding operations must be performed sufficiently 
fast to maintain coherent rotations within the nearly degenerate ground-state manifold~\cite{Harper,Amorim2015}.
Interestingly, we find that at $t=0$, the ground-state energy splitting $\delta E$ is equal in magnitude 
to the Coulomb interaction (i.e., $\delta E = V$). 
This splitting, induced by the Coulomb repulsion $V$, sets an upper bound on the 
allowable switching time $\tau$, approximately given by $\tau \lesssim \hbar / \delta E$. 
The switching time $\tau$ must be sufficiently small so that the braiding operation is fast enough 
to coherently mix the split ground states~\cite{Peeters2024,Hodge2025PRL,Knapp2016}. 
Combining both the lower bound (set by diabatic errors) and the upper bound (set by Coulomb interaction $V$) 
yields the constraint on switching time
  $\frac{\hbar}{\Delta_{\text{gap}}} \ \lesssim \ \tau \ \lesssim \ \frac{\hbar}{\delta E}$.\\

\noindent {\small \bf Initialization and readout of the MZM-qubit}

\begin{figure*}
\hspace*{-0.5cm}
\vspace*{0cm}
\begin{overpic}[width=2.1\columnwidth]{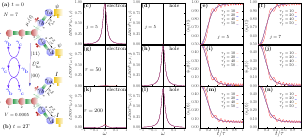}
\end{overpic}
\caption{{\bf Initialization and readout of a Majorana qubit via $Z$-fusion in an $X$-gate protocol.}
\textbf{a}, At $t = 0$, the system is initialized with two spatially separated pairs of Majorana zero modes (MZMs),
$(\gamma_a, \gamma_b)$ and $(\gamma_c, \gamma_d)$. $Z$-fusion operations $F^Z_{ba}$ and $F^Z_{cd}$ 
result in the formation of two electrons $\psi$ localized at sites $j=5$ and $j=7$, corresponding to the qubit state $|11\rangle$.
\textbf{b}, After a double braiding of $(\gamma_b, \gamma_c)$ at $t = 2T$, the $Z$-fusion operations are repeated, yielding vacuum states ($I$) at both sites, 
corresponding to the qubit state $|00\rangle$.
\textbf{c--f}, Fusion readout without braiding for various fusion switching times $\tau_f$. 
\textbf{c}, Electron component of the local density of states, $LDOS^e(\omega, t, j)$, at site $j=5$.                                                                
\textbf{d}, Hole component, $LDOS^h(\omega, t, j)$, at site $j=5$.
\textbf{e}, Time-dependent charge density $\langle n_j(t) \rangle$ at site $j=5$.                     
\textbf{f}, Time-dependent charge density at site $j=7$.
\textbf{g--j}, Readout after double braiding for $\tau = 50$. 
\textbf{g}, $LDOS^e(\omega, t, j)$ versus $\omega$ at $j=5$.
\textbf{h}, $LDOS^h(\omega, t, j)$ versus $\omega$ at $j=5$.                                              
\textbf{i,j}, Time-dependent charge densities $\langle n_j(t) \rangle$ at $j=5$ and $j=7$.
\textbf{k--n}, Readout after longer braid duration $\tau = 200$.                               
\textbf{k}, $LDOS^e(\omega, t, j)$ versus $\omega$ at $j=5$.                                                 
\textbf{l}, $LDOS^h(\omega, t, j)$ versus $\omega$ at $j=5$.                                 
\textbf{m,n}, Time-dependent charge densities $\langle n_j(t) \rangle$ at sites $j=5$ and $j=7$.}
\label{fig4}
\end{figure*}

We employ a simple $Z$-fusion of Majorana zero modes (MZMs), 
which is fusion within the same MZM pair~\cite{Pandey2023_PRB,Hodge2025}, 
as a tool for both initialization and readout of the MZM-based qubit.
MZMs obey the fusion rule $\gamma \times \gamma = I + \psi$, where $I$ denotes the vacuum and $\psi$ 
corresponds to a fermionic state (electron)~\cite{Nayak}. 
Experimentally, the outcome of such fusion can be accessed either by resolving 
the electron and hole components of the local density of states (LDOS)~\cite{Jack2021}, 
or by measuring the localized charge after fusion~\cite{Szechenyi2020,Gharavi2016}. 
In our setup (Fig.~\ref{fig4}(a)), the system is initialized at time $t = 0$ with two separate MZM pairs, 
$(\gamma_a, \gamma_b)$ and $(\gamma_c, \gamma_d)$, both occupying parity sectors $P_{ba} = P_{cd} = -1$. 
To perform the \( Z \)-fusions \( F^Z_{ba} \) and \( F^Z_{cd} \), the on-site chemical potentials \( \mu_j(t) \) at sites 
\( j = 4 \) and \( j = 5 \) are increased simultaneously.

In Figs.~\ref{fig4}(c,d), we present the normalized electron and hole components of the local density of states 
($LDOS(\omega,t,j)$) at site \( j = 5 \) for different fusion switching times \( \tau_f \), prior to performing the braiding of MZMs. 
At the final fusion time \( t = t_f \), and for \( \tau_f \sim 50 \), the electron \( \text{LDOS}^e(\omega, t, j) \), 
at site \( j = 5 \) (and similarly at \( j = 7 \), not shown) 
exhibits a zero-energy peak of unit height (Fig.~\ref{fig4}(c)), 
while the hole component shows negligible spectral weight (Fig.~\ref{fig4}(d)). 
This indicates the successful formation of an electron \( \psi \) on site \( j = 5 \) following the fusion process.
We also compute the time-dependent local charge density \( \langle n_j(t) \rangle \) at sites \( j = 5 \) and \( j = 7 \) 
[Figs.~\ref{fig4}(e,f)]. For small fusion switching times \( \tau_f \), 
the charge density exhibits oscillations due to non-adiabatic excitations induced by 
rapid variations in \( \mu_j(t) \). However, for \( \tau_f \gtrsim 40 \), 
the charge density at both sites approaches unity, indicating the formation of an electron \( \psi \) on each site.
This readout process, implemented via simple \( Z \)-fusion, provides a clear signature of the initial occupancy of 
the MZM qubit ($|11\rangle$) formed by the two MZM pairs prior to braiding.

Next, we perform a double braiding operation $U^2_{bc}$ between MZMs $\gamma_b$ and $\gamma_c$ (see schematic of Fig.~\ref{fig4}(b)).
As discussed earlier, this operation transforms the MZM qubit state from $|11\rangle$ to $|00\rangle$.
To read out the final state after double braiding (at $t = 2T$), we repeat the same $Z$-fusions $F^Z_{ba}$ and $F^Z_{cd}$.
The second row of Figs.~\ref{fig4}(g-j) presents the results of $Z$-fusion for different fusion switching times $\tau_f$,
following a braiding protocol with $\tau = 50$ (braiding switching time).
In this case, the hole component of the local density of states, $LDOS^h(\omega, t, j)$, 
exhibits a peak with height one at $\omega = 0$ (Fig.~\ref{fig4}(h)), while the electron component, $LDOS^e(\omega, t, j)$, 
shows no peak at site $j = 5$ (Fig.~\ref{fig4}(g)). This indicates the formation of a vacuum state ($|00\rangle$) after $Z$-fusion, 
in contrast to the electron formation $|11\rangle$ observed prior to braiding.
Figures~\ref{fig4}(i,j), show the time-dependent charge density $\langle n_j(t) \rangle$ at sites $j = 5$ and $j = 7$, respectively, 
for various $\tau_f$. In both cases, the charge density approaches zero for $\tau_f \gtrsim 40$, 
confirming the formation of holes at both sites following fusion.
These results validate that the double braid operation leads to a state transformation $|11\rangle \rightarrow |00\rangle$ and 
should be obtained in the quantum-dot experiment using charge sensor~\cite{Szechenyi2020,Gharavi2016}.

The third row, Figs.~\ref{fig4}(k–n), presents the results of \( Z \)-fusion following a braiding protocol with a 
longer switching time $\tau = 200$ and $V=0.005$, for which the braiding fidelity decreases to approximately 0.93 (see Fig.~\ref{fig3}(j)).
In this case, the hole component of the local density of states, \( \text{LDOS}^h(\omega, t, j) \), shows a peak at \( \omega = 0 \) 
with a reduced height of approximately 0.93 [Fig.~\ref{fig4}(l)], while the electron component, \( \text{LDOS}^e(\omega, t, j) \), 
shows a residual peak of height ~0.07 [Fig.~\ref{fig4}(k)].
Correspondingly, the time-dependent charge densities \( \langle n_j(t) \rangle \) at sites 
\( j = 5 \) [Fig.~\ref{fig4}(m)] and \( j = 7 \) [Fig.~\ref{fig4}(n)] saturate around 0.07, 
indicating the presence of finite errors in the final state. 
These results demonstrate that the decrease in braiding fidelity for longer braiding durations 
is reflected in the charge readout errors obtained through fusion of the MZM-based qubit.\\

\begin{figure}
\centering
\includegraphics[width=0.48\textwidth]{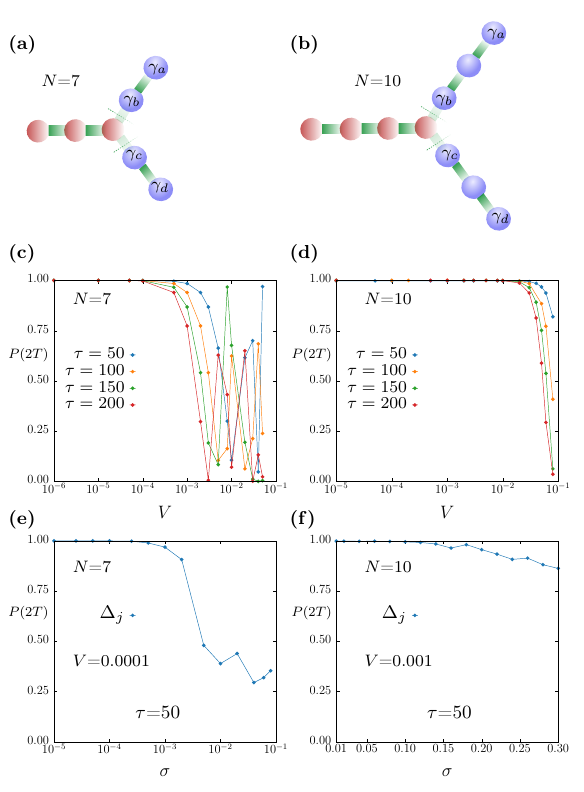}
\caption{{\bf Topological protection and braiding fidelity.}
\textbf{a}, Schematic of a $Y$-shaped geometry with $N = 7$ quantum dots hosting two pairs of Majorana zero modes (MZMs), 
$(\gamma_a, \gamma_b)$ and $(\gamma_c, \gamma_d)$, localized at sites $j = 4,5$ and $j = 6,7$, respectively. 
\textbf{b}, Schematic of a $Y$-shaped geometry with $N = 10$ quantum dots, where the MZM pairs 
$(\gamma_a, \gamma_b)$ and $(\gamma_c, \gamma_d)$ are localized at sites $j = 5,7$ and $j = 8,10$, respectively. 
Braiding fidelity $P(2T)$ as a function of nearest-neighbor Coulomb interaction strength $V$ (logarithmic scale) 
is shown for different switching times $\tau$ with \textbf{c}, $N = 7$, and \textbf{d}, $N = 10$. 
Braiding fidelity $P(2T)$ as a function of disorder strength $\sigma$ is shown for 
\textbf{e}, $N = 7$, with $V = 0.0001$ and $\tau = 50$ (logarithmic scale), and 
\textbf{f}, $N = 10$, with $V = 0.001$ and $\tau = 50$ (linear scale in $\sigma$).}
\label{fig5}
\end{figure}

\noindent {\small \bf Topological protection against interactions and disorder.} 
The braiding of Majorana zero modes (MZMs) is topologically protected as long as the MZMs remain spatially 
well-separated and do not overlap~\cite{Sarma,Peeters2024}. This topological protection provides inherent resilience against various 
imperfections present in quantum-dot-based systems. In such systems, the dominant sources of imperfection include nearest-neighbor 
Coulomb interactions and disorder in the pairing amplitudes, which can shift the system away from the sweet spot~\cite{Dvir,Bordin}.
In hybrid semiconductor–superconductor quantum-dot platforms, Coulomb interactions can be effectively screened and minimized by the central 
superconducting island, which reduces interdot coupling~\cite{Tsintzis2024}. 
In Figs.~\ref{fig5}(c,d), we investigate the effect of repulsive nearest-neighbor Coulomb interactions \( V \) on the braiding fidelity \( P(2T) \) 
for two system sizes, \( N = 7 \) and \( N = 10 \), and for different values of the switching time \( \tau \). 
We find that for fast braiding protocols (e.g., \( \tau \sim 50 \)), the fidelity remains close to unity up to 
interaction strengths of \( V \sim 0.001 \) [Fig.~\ref{fig5}(c)]. However, as \( V \) increases, hybridization between 
MZMs becomes more pronounced, resulting in strong oscillations and a significant drop in braiding fidelity. 
Interestingly, the larger system with \( N = 10 \) exhibits enhanced robustness: 
the fidelity remains near unity even at comparatively strong interaction strengths of \( V \sim 0.04 \) [Fig.~\ref{fig5}(d)].

To understand the effect of disorder on the braiding of Majorana zero modes (MZMs), 
we introduce site-dependent disorder in the superconducting pairing amplitude \( \Delta_k \). 
Each disordered component is independently drawn from a normal distribution:
\begin{equation}
    g(x) = \frac{1}{\sigma \sqrt{2\pi}} e^{ -\frac{1}{2} \left( \frac{x - a}{\sigma} \right)^2 },
\end{equation}
where \( \sigma \) is the standard deviation and \( a \) is the mean of the distribution. 
For disorder in \( \Delta_k \), we set \( a = 1 \), corresponding to the clean-limit value of the pairing amplitude.
In Fig.~\ref{fig5}(e), we present the disorder-averaged braiding fidelity \( P(2T) \), computed over sixty disorder realizations,
as a function of disorder strength \( \sigma \) in the pairing amplitudes \( \Delta_j \), for a system of size \( N = 7 \) with interaction strength \( V = 0.0001 \). We observe that as the disorder strength increases, the braiding fidelity \( P(2T) \) drops below 0.99 for \( \sigma \gtrsim 0.0005 \).
This reduction in fidelity arises because increasing disorder enhances the localization length of the MZMs, leading to their hybridization and 
degradation, particularly for those MZMs residing on nearest-neighbor sites. Interestingly, as shown in Fig.~\ref{fig5}(f), for a larger system with \( N = 10 \), the fidelity remains above 0.99 up to disorder strengths of \( \sigma \sim 0.12 \), indicating improved robustness against pairing disorder.

The enhanced topological protection against Coulomb repulsion and disorder in the pairing amplitude is attributed to the increased spatial 
separation of MZM pairs in the \( N = 10 \) quantum-dot setup, where the MZMs reside on three distinct sites, 
thereby reducing direct overlap~\cite{Bordin}. 
In contrast, for \( N = 7 \), the initial MZMs occupy neighboring sites, making them more susceptible to hybridization~\cite{Dvir}. 
Furthermore, the larger system size suppresses MZM overlap throughout the braiding process, thereby enhancing topological 
protection even in the presence of interactions and pairing disorder. Thus, one of our primary theoretical
predictions is that the larger the number of quantum dots used experimentally, the better the expected experimental outcome regarding simulating quantum computing operations.

%%%%%%%%%%%%%%%%

\noindent \textbf{Discussion} \\
In this work, we have shown the non-Abelian statistics of Majorana zero modes (MZMs) 
and implemented the Pauli $X$-quantum gate using only a few quantum dots arranged in a $Y$-shaped geometry.
By employing a time-dependent many-body exact diagonalization method, we have computed diabatic errors, 
braiding fidelity, and the time-, space-, and energy-resolved local density of states ($LDOS(\omega, j, t)$)
to analyze the dynamics of the braiding protocol in a quantum-dot-based system. The time-dependent
 $LDOS(\omega, j, t)$ which can be experimentally accessed using scanning tunneling spectroscopy (STS), provides a direct visualization of the spatial location of MZMs throughout the braiding process. Our results reveal that the MZMs in the $Y$-geometry are spatially extended over multiple quantum dots, 
and indicate that a minimum of seven quantum dots is necessary to successfully complete the braiding operation without unwanted overlap between MZMs.

For the seven–quantum-dot system, we find that nearest-neighbor Coulomb interactions are highly detrimental to the braiding operation, 
leading to pronounced oscillations in the braiding fidelity. This behavior imposes the upper bounds on the switching time \(\tau\), 
which governs the braiding speed in interacting quantum-dot systems~\cite{Peeters2024,Hodge2025PRL,Knapp2016}. 
Our numerical simulations reveal that the braiding protocol achieves optimal fidelity for switching times in the range \(\tau \in [40,100]\). 
For \(\tau=50\), and using the relation \(t_{\mathrm{phys}}=(12\tau)\,\hbar/\Delta_{\mathrm{gap}}\) for a full double-braiding operation, 
together with a typical effective superconducting gap \(\Delta_{\mathrm{gap}} \in [25~\mu\mathrm{eV},\,75~\mu\mathrm{eV}]\)~\cite{Dvir,Bordin}, 
we estimate a total duration of approximately \(5.27~\mathrm{ns}\) to \(15.80~\mathrm{ns}\). 
This provides the physical timescale associated with the lower bound on the braid duration.
For \(N=7\) quantum dots, we find that high braiding fidelity is achievable for \(V\!\lesssim\!0.001\).
For \(V=0.001\), the splitting of the ground-state degeneracy is approximately \(\delta E \lesssim 0.001\,\Delta\), 
which induces oscillations in the braiding fidelity and sets an upper bound on the switching time. 
In SI units, using \(\Delta_{\mathrm{gap}} \in [25,75]~\mu\mathrm{eV}\) and the relation \(t_{\text{upper}} = (12\,\hbar)/V\), 
we obtain \(t_{\text{upper}} \in [105~\mathrm{ns},\,316~\mathrm{ns}]\). Combining these upper and lower bounds, 
the operational window for successful braiding at \(V=0.001\) is therefore
 $ 5.3~\mathrm{ns} \;\lesssim\; t_{\text{braid}} \;\lesssim\; 316~\mathrm{ns}$,
which is both fast enough to avoid interaction-induced oscillations and slow enough to suppress diabatic errors.
Notably, the coherence time of MZM-based qubits in quantum-dot systems has been estimated to be on the order of \(\sim\!1~\mathrm{ms}\)~\cite{Bordin}, 
several orders of magnitude longer than the timescale required for implementing a full braiding operation or executing the Pauli-\(X\) gate. 
These findings indicate that the braiding operations proposed here can, in principle, be carried out well within the coherence time of the 
quantum-dot platform, even when limited to seven quantum dots~\cite{Bordin}.

Next, we show that a simple $Z$-fusion operation within the same pair of MZMs can be 
employed to both initialize the system (prior to braiding) and read out the final state (after braiding) in quantum-dot-based experiments. Our numerical calculations, based on the time-dependent charge density and the electron and hole components of the local density of states, $LDOS(\omega, j, t)$, 
reveal distinct and opposite fusion outcomes before and after the braiding process. 
This contrast in fusion outcomes provides a direct and experimentally accessible signature of the non-Abelian statistics of MZMs. In quantum dot experiments, such fusion outcomes can be detected using local charge-sensing probes.

Finally, we examined the topological protection of the braiding protocol against nearest-neighbor 
Coulomb interactions and disorder in the pairing amplitude. 
By compairing the braiding fidelity for two system sizes, $N = 7$ and $N = 10$, 
we find that topological protection improves significantly with increasing system size. In particular, 
the $N = 10$ quantum dot system is more robust against both interactions and pairing disorder. 
We also analyzed the effect of disorder in the on-site chemical potential (see Supplementary Material), 
where we observe that for $N = 7$ quantum dots, the braiding fidelity survives at higher disorder strengths compared to the case with pairing disorder. However, on-site disorder can directly affect the accuracy of local charge measurements after the fusion of MZMs.
Fortunately, quantum dot systems offer highly tunable control over the 
local chemical potential of individual dots via gate voltages. 
We believe that our proposed architecture, using seven quantum dots for braiding and 
implementing the $X$-gate in a $Y$-shaped geometry, can be realized in recently developed quantum 
dot platforms. Our proposed charge-based detection method, 
relying on time-dependent $LDOS(\omega,j,t)$ and local charge density after MZM fusion, 
offers a viable and accessible route for observing non-Abelian statistics in current experimental setups.

\noindent {\small \bf Method\\} 
\noindent {\small \bf Local density-of-states\\}
We have calculated the time-dependent local density of states 
using the exact diagonalization method~\cite{Pandey2023_PRB}. 
To simulate the time evolution, we propagate the initial wavefunction $|\Psi(0)\rangle$
using the time-dependent Hamiltonian $H(t)$ up to time $t$:
\begin{equation}
|\Psi(t)\rangle = \mathcal{T} \exp\left( -i \int_0^t H(s)\, ds \right) |\Psi(0)\rangle,
\end{equation}
where $\mathcal{T}$ denotes the time-ordering operator~\cite{Kells2014}. 
We then evaluate the double-time Green's function~\cite{Kennes2014} $G(t,t')$ using the final Hamiltonian $H_f = H(t_f)$
at the final time $t = t_f$. For the electronic component, the Green's function is given by:
\begin{equation}
G^{\text{elec}}_j(t,t') = \langle \Psi(t) | c_j^\dagger\, e^{i H_f t'}\, c_j\, e^{-i H_f t'} | \Psi(t) \rangle.
\end{equation}

The electronic part of the time-dependent local density of states at site $j$ is obtained by performing a 
Fourier transform of the Green's function $G^{\text{elec}}_j(t,t')$ with respect to $t'$:

\begin{equation}
\text{LDOS}^{\text{e}}_j(\omega, t) = \frac{1}{\pi} \, \text{Im} \left[ \int_0^T dt'\, e^{i (\omega + i \eta) t'}\, i G^{\text{elec}}_j(t, t') \right],
\end{equation}
where the integration is carried out up to $T = 60$, and the energy broadening parameter is set to $\eta = 0.1$ throughout this work.

In an analogous way, the hole component of the LDOS is derived from the corresponding hole Green's function:
\begin{equation}
G^{\text{hole}}_j(t,t') = \langle \Psi(t) | c_j(t')\, c_j^\dagger | \Psi(t) \rangle,
\end{equation}
The total local density of states at site $j$ is given by the sum of the electron and hole components:
\begin{equation}
\text{LDOS}_j(\omega, t) = \text{LDOS}^{\text{e}}_j(\omega, t) + \text{LDOS}^{\text{h}}_j(\omega, t).
\end{equation}

\noindent {\bf {\small \\ Data availability\\}} The data that support the findings of this study are available from the corresponding author upon request.

\noindent {\bf {\small \\ Code availability\\}} 
The computer codes used in this study are available from the
corresponding author upon request.

\noindent {\bf {\small \\Funding Declaration\\}}
The work of B.P., S.O., and E.D. was supported by the
U.S. Department of Energy (DOE), Office of Science, Basic
Energy Sciences (BES), Materials Sciences and Engineering
Division. S.R.\ acknowledges support from the Australian Research Council through Grant No.\ DP240100168.

\noindent {\bf {\small \\ Author contributions\\}}
B.P. designed the project and carried out the numerical calculations.
B.P. and E.D. wrote the manuscript.
S.O., P.L., and S.L. provided valuable comments and discussions.
All authors reviewed the manuscript.

\noindent {\bf {\small \\ Competing interests\\}} The authors declare no competing interests.

\noindent {\bf {\small \\ Additional information\\}}
Correspondence and requests for materials should be addressed to  Bradraj Pandey ({\it bradraj.pandey@gmail.com}).

%\begin{thebibliography}{10}
%\bibliographystyle{ieeetr} 

%-----------------------------------------------------------------


\begin{thebibliography}{99}
\bibitem{Kitaev1}Kitaev AY. {Unpaired Majorana fermions in quantum wires.}
\href{https://doi.org/10.1070/1063-7869/44/10S/S29}{\it Phys.-Usp.} {\textbf{44}, 131 (2001)}.

\bibitem{Kitaev2}Kitaev AY. {Fault-tolerant quantum computation by anyons.}
\href{https://doi.org/10.1016/S0003-4916(02)00018-0}{\it Ann Phys (NY)} {\textbf{303}, 2 (2003)}.

\bibitem{Sarma}Sarma, S., Freedman, M.  Nayak, C. {Majorana zero modes and topological quantum computation.}
\href{https://doi.org/10.1038/npjqi.2015.1}{\it  npj Quantum Inf } {\textbf{1}, 15001 (2015)}.


\bibitem{Nayak}Nayak C, Simon SH, Stern A, Freedman M,  Sarma SD. {Non-abelian anyons and topological quantum computation.}
\href{https://doi.org/10.1103/RevModPhys.80.1083}{\it Rev Mod Phys } {\textbf{80}, 1083 (2008)}.


\bibitem{Shnirman} Scheurer, M. S., \& Shnirman. {Nonadiabatic processes in Majorana qubit systems.}
\href{https://doi.org/10.1103/PhysRevB.88.064515} {\it Phys. Rev. B} {\textbf{88}, 064515 (2013)}.



\bibitem{Sau}Lutchyn, R. M., Sau, J. D., Das Sarma S. {Majorana fermions and a topological phase transition in semiconductor–superconductor heterostructures.}
\href{https://doi.org/10.1038/ncomms1966}{\it Phys. Rev. Lett.} {\textbf{105}, 077001 (2010)}.

\bibitem{Oreg2010}
Oreg Y, Refael G, von Oppen F. {Helical liquids and Majorana bound states in quantum wires.}
\href{https://doi.org/10.1103/PhysRevLett.105.177002}{\textit{Phys Rev Lett} \textbf{105}, 177002 (2010)}.


\bibitem{Yazdani}
	Nadj-Perge, S., Drozdov, I. K., Li, J., Chen, H., Jeon, S., Seo, J., MacDonald, A. H., Bernevig, B. A. \& Yazdani, A. {Observation of Majorana fermions in ferromagnetic atomic chains on a superconductor.}
\href{https://doi.org/10.1126/science.1259327}{\textit{Science} \textbf{346}, 602--607 (2014)}.


\bibitem{Jay}Sau, J., Sarma, S. {Realizing a robust practical Majorana chain in a quantum-dot-superconductor linear array.}
\href{https://doi.org/10.1038/ncomms1966}{\it  Nat Commun.} {\textbf{3}, 964 (2012)}.

\bibitem{Dvir}
	Dvir, T., Wang, G., van Loo, N., Liu, C.-X., Mazur, G. \textit{et al.} {Realization of a minimal Kitaev chain in coupled quantum dots.} \href{https://doi.org/10.1038/s41586-023-06025-y}{\textit{Nature} \textbf{614}, 445–450 (2023)}.

\bibitem{Bordin}
Bordin, A., Liu, C.-X., Dvir, T. \textit{et al.} Enhanced Majorana stability in a three-site Kitaev chain. 
\href{https://doi.org/10.1038/s41565-025-01894-4}{\textit{Nat. Nanotechnol.} \textbf{20}, 726--731 (2025)}.


\bibitem{tenHaaf}
ten Haaf, S. L. D., Zhang, Y., Wang, Q. \textit{et al.} 
{Observation of edge and bulk states in a three-site Kitaev chain.} 
\href{https://doi.org/10.1038/s41586-025-08892-5}{\textit{Nature} \textbf{641}, 890--895 (2025)}.


\bibitem{Boross2024}
Boross, P. \& Pályi, A. Braiding-based quantum control of a Majorana qubit built from quantum dots. 
\href{https://doi.org/10.1103/PhysRevB.109.125410}{\textit{Phys. Rev. B} \textbf{109}, 125410 (2024)}.

\bibitem{Tsintzis2024}
Tsintzis, A., Seoane Souto, R., Flensberg, K. \textit{et al.} Majorana qubits and non-Abelian physics in quantum dot--based minimal Kitaev chains.
\href{https://doi.org/10.1103/PRXQuantum.5.010323}{\textit{PRX Quantum} \textbf{5}, 010323 (2024)}.

\bibitem{Miles2025}
Miles, S., Zatelli, F., Bozkurt, A. M. \textit{et al.} Braiding Majoranas in a linear quantum dot–superconductor array: Mitigating the errors from Coulomb repulsion and residual tunneling. 
\href{https://doi.org/10.48550/arXiv.2501.16056}{arXiv:2501.16056 (2025)}.

\bibitem{Sekania2017}
Sekania M, Plugge S, Greiter M, Thomale R, Schmitteckert P. {Braiding errors in interacting Majorana quantum wires.}
\href{https://doi.org/10.1103/PhysRevB.96.094307}{\it Phys Rev B} {\bf 96}, 094307 (2017).

\bibitem{Pandey2023}
Pandey, B., Kaushal, N., Alvarez, G. \textit{et al.} Majorana zero modes in Y-shape interacting Kitaev wires. 
\href{https://doi.org/10.1038/s41535-023-00584-5}{\textit{npj Quantum Mater.} \textbf{8}, 51 (2023)}.

\bibitem{Pandey2025}
Pandey, B., Alvarez, G., Dagotto, E. \& Zhang, R.-X. Crystalline-symmetry-protected Majorana modes in coupled quantum dots.
\href{https://doi.org/10.1103/PhysRevResearch.7.L012022}{\textit{Phys. Rev. Res.} \textbf{7}, L012022 (2025)}.


\bibitem{Pandey2025_PRB}
Pandey, B., Gupta, G. K., Alvarez, G. \textit{et al.} Diabatic error and propagation of Majorana zero modes in interacting quantum dots systems.
\href{https://doi.org/10.1103/PhysRevB.111.104311}{\textit{Phys. Rev. B} \textbf{111}, 104311 (2025)}.



\bibitem{Harper}
Harper, F., Pushp, A. \& Roy, R. Majorana braiding in realistic nanowire Y-junctions and tuning forks. 
\href{https://doi.org/10.1103/PhysRevResearch.1.033207}{\textit{Phys. Rev. Res.} \textbf{1}, 033207 (2019)}.

\bibitem{Amorim2015}
Amorim, C. S., Ebihara, K., Yamakage, A. \textit{et al.} Majorana braiding dynamics in nanowires. 
\href{https://doi.org/10.1103/PhysRevB.91.174305}{\textit{Phys. Rev. B} \textbf{91}, 174305 (2015)}.







\bibitem{Pandey2023_PRB}
Pandey, B., Mohanta, N. \& Dagotto, E. Out-of-equilibrium Majorana zero modes in interacting Kitaev chains. 
\href{https://doi.org/10.1103/PhysRevB.107.L060304}{\textit{Phys. Rev. B} \textbf{107}, L060304 (2023)}.

\bibitem{Thomale2013}
Thomale, R., Rachel, S. \& Schmitteckert, P. Tunneling spectra simulation of interacting Majorana wires. 
\textit{Phys. Rev. B} \textbf{88}, 161103 (2013). 
\href{https://doi.org/10.1103/PhysRevB.88.161103}{https://doi.org/10.1103/PhysRevB.88.161103}.


\bibitem{Houselt2010}
van Houselt A, Zandvliet HJW. {Colloquium: time-resolved scanning tunneling microscopy.}
\href{https://doi.org/10.1103/RevModPhys.82.1593}{\it Rev Mod Phys} {\bf 82}, 1593–1605 (2010).

\bibitem{Szechenyi2020}
Széchenyi, G. \& Pályi, A. Parity-to-charge conversion for readout of topological Majorana qubits.
\href{https://doi.org/10.1103/PhysRevB.101.235441}{\textit{Phys. Rev. B} \textbf{101}, 235441 (2020)}.






\bibitem{Gharavi2016}
Gharavi, K., Hoving, D. \& Baugh, J. Readout of Majorana parity states using a quantum dot.
\href{https://doi.org/10.1103/PhysRevB.94.155417}{\textit{Phys. Rev. B} \textbf{94}, 155417 (2016)}.


\bibitem{Bordin2023}
Bordin, A., Wang, G., Liu, C.-X. \textit{et al.} Tunable crossed Andreev reflection and elastic cotunneling in hybrid nanowires.
\href{https://doi.org/10.1103/PhysRevX.13.031031}{\textit{Phys. Rev. X} \textbf{13}, 031031 (2023)}.

\bibitem{Liu2025}
Liu, C.-X., Miles, S., Bordin, A. \textit{et al.} Scaling up a sign-ordered Kitaev chain without magnetic flux control. 
\href{https://doi.org/10.1103/PhysRevResearch.7.L012045}{\textit{Phys. Rev. Res.} \textbf{7}, L012045 (2025)}.

\bibitem{Pandey2024_PRR}
Pandey, B., Okamoto, S. \& Dagotto, E. Nontrivial fusion of Majorana zero modes in interacting quantum-dot arrays.
\href{https://doi.org/10.1103/PhysRevResearch.6.033314}{\textit{Phys. Rev. Res.} \textbf{6}, 033314 (2024)}.

\bibitem{Marra2022}
Marra, P. Majorana nanowires for topological quantum computation. 
\href{https://doi.org/10.1063/5.0102999}{\textit{J. Appl. Phys.} \textbf{132}, 231101 (2022)}.


\bibitem{Malciu2018}
Malciu, C., Mazza, L. \& Mora, C. Braiding Majorana zero modes using quantum dots.
\href{https://doi.org/10.1103/PhysRevB.98.165426}{\textit{Phys. Rev. B} \textbf{98}, 165426 (2018)}.

\bibitem{Hodge2025}
Hodge, T., Kieu, T., Bedow, J. \textit{et al.} Fusion dynamics of Majorana zero modes. 
\href{https://doi.org/10.48550/arXiv.2503.09800}{arXiv:2503.09800 (2025)}.


\bibitem{Clarke2017}
Clarke, D. J., Sau, J. D. \& Das Sarma, S. Probability and braiding statistics in Majorana nanowires. 
\href{https://doi.org/10.1103/PhysRevB.95.155451}{\textit{Phys. Rev. B} \textbf{95}, 155451 (2017)}.

\bibitem{Mascot2023}
Mascot, E., Hodge, T., Crawford, D. \textit{et al.} Many-body Majorana braiding without an exponential Hilbert space. 
\href{https://doi.org/10.1103/PhysRevLett.131.176601}{\textit{Phys. Rev. Lett.} \textbf{131}, 176601 (2023)}.


\bibitem{Bedow2024}
Bedow, J., Mascot, E., Hodge, T. \textit{et al.} Simulating topological quantum gates in two-dimensional magnet–superconductor hybrid structures.
\href{https://doi.org/10.1038/s41535-024-00703-w}{\textit{npj Quantum Mater.} \textbf{9}, 99 (2024)}.



\bibitem{Knapp2016}
Knapp, C., Zaletel, M., Liu, D. E. \textit{et al.} The nature and correction of diabatic errors in anyon braiding.
\href{https://doi.org/10.1103/PhysRevX.6.041003}{\textit{Phys. Rev. X} \textbf{6}, 041003 (2016)}.

\bibitem{Peeters2024}
Peeters, C., Hodge, T., Mascot, E. \textit{et al.} Effect of impurities and disorder on the braiding dynamics of Majorana zero modes.
\href{https://doi.org/10.1103/PhysRevB.110.214506}{\textit{Phys. Rev. B} \textbf{110}, 214506 (2024)}.

\bibitem{Hodge2025PRL}
Hodge, T., Mascot, E., Crawford, D. \textit{et al.} Characterizing dynamic hybridization of Majorana zero modes for universal quantum computing. 
\href{https://doi.org/10.1103/PhysRevLett.134.096601}{\textit{Phys. Rev. Lett.} \textbf{134}, 096601 (2025)}.

\bibitem{Jack2021}
Jäck, B., Xie, Y. \& Yazdani, A. Detecting and distinguishing Majorana zero modes with the scanning tunnelling microscope. 
\href{https://doi.org/10.1038/s42254-021-00328-z}{\textit{Nat. Rev. Phys.} \textbf{3}, 541--554 (2021)}.


\bibitem{Kells2014}Kells G, Sen D, Slingerland JK, Vishveshwara S. {Topological degeneracy and vortex manipulation in Kitaev's honeycomb model.}
\href{https://doi.org/10.1103/PhysRevB.89.235130}{\it Phys Rev B } {\textbf{89}, 235130 (2014)}.

\bibitem{Kennes2014}Kennes DM, Klöckner C, Meden V. {Nonadiabatic dynamics of interacting quantum dots: Renormalization group approach.}
\href{https://doi.org/10.1103/PhysRevLett.113.116401}{\it Phys Rev Lett } {\textbf{113}, 116401 (2014)}.

\end{thebibliography}
\end{document}